\documentclass[sigplan,screen]{acmart}
\usepackage[T1]{fontenc}
\usepackage[utf8]{inputenc}
\usepackage{microtype}
\usepackage{amsmath}
\usepackage{graphicx}
\usepackage{subcaption}
\usepackage{booktabs}
\usepackage{algorithm}
\usepackage{algpseudocode}
\usepackage{enumitem}

\acmYear{2026}\copyrightyear{2026}
\setcopyright{cc}
\setcctype[4.0]{by}
\acmConference[EuroMLSys '26]{The 6th Workshop on Machine Learning and Systems}{April 27--30, 2026}{Edinburgh, Scotland Uk}
\acmBooktitle{The 6th Workshop on Machine Learning and Systems (EuroMLSys '26), April 27--30, 2026, Edinburgh, Scotland Uk}
\acmDOI{10.1145/3805621.3807652}
\acmISBN{979-8-4007-2605-7/26/04}


\title{Accuracy Is Speed: Towards Long-Context-Aware Routing for Distributed LLM Serving}

\author{Takeshi Yoshimura}
\authornote{These authors contributed equally to this work.}
\affiliation{\institution{IBM Research - Tokyo}\city{Tokyo}\country{Japan}}
\email{tyos@jp.ibm.com}
\author{Valentijn Dymphnus van de Beek}
\authornotemark[1]
\affiliation{\institution{Delft University of Technology}\city{Delft}\country{Netherlands}}
\email{v.d.van.de.beek@student.tudelft.nl}
\author{Tatsuhiro Chiba}
\affiliation{\institution{IBM Research - Tokyo}\city{Tokyo}\country{Japan}}
\email{chiba@jp.ibm.com}

\begin{document}

\begin{abstract}

Distributed LLM serving systems optimize per-request latency and throughput.
However, under long-context workloads, inference accuracy becomes more variable.
When incorrect responses trigger retries, accuracy directly translates into cumulative user-visible delay that is not captured by single-shot latency metrics.

In this work, we argue that under long-context serving, \textbf{accuracy becomes speed} through retry dynamics.
We introduce \textit{Time-to-Correct-Answer (TTCA)}, a metric that measures the wall-clock time required to obtain the first correct response.
Our measurement study shows that prompt characteristics such as length and language amplify accuracy variance, which inflates TTCA.
We demonstrate \textit{Lightweight Accuracy-Aware Routing (LAAR)}, a capability-based routing design that reduces TTCA.
Our results suggest that in long-context distributed serving, accuracy should be treated as a first-class systems objective.

\end{abstract}


\maketitle

\section{Introduction}

Large language models (LLMs) are increasingly used in long-context settings, such as retrieval-augmented generation \cite{jin2024longcontextllmsmeetrag}, code assistants \cite{haseeb2025contextengineeringmultiagentllm}, tool-using agents \cite{xu2026evolutiontoolusellm}, and document understanding \cite{10.5555/3737916.3740957}.
Recent systems work demonstrates that serving long prompts with hundreds of thousands or even millions of tokens is now technically feasible \cite{contextparallel2025,yang2025lserveefficientlongsequencellm}.

Long-context workloads fundamentally change the performance characteristics of LLM inference.
Prefill computation dominates cost \cite{yang2025lserveefficientlongsequencellm}, memory bandwidth becomes a primary bottleneck \cite{11120553}, and cache management plays a critical role in overall efficiency \cite{vllm_sosp23}.
These factors make per-request load and cache locality more important in long-context serving.
Distributed LLM serving systems have introduced load-aware, session-affinity, and cache-affinity routing strategies to optimize latency and throughput \cite{llmd-inference-scheduler,yuan2026dualmap,performanceaware2025}.

However, long-context workloads affect not only performance but also accuracy.
Our measurements show that accuracy varies with prompt length, model, and language under long inputs.
This variation complicates routing among heterogeneous instances in a cluster.
When incorrect answers trigger retries or escalation to larger models, routing mistakes inflate cumulative end-to-end latency.
In long-context distributed serving, \emph{accuracy therefore becomes a system-level performance factor}, and one-shot latency alone no longer captures user-visible delay.
In this sense, higher accuracy can directly reduce end-to-end latency.
Accuracy therefore effectively becomes a form of speed.

To capture this effect, we introduce \emph{Time-to-Correct-Answer (TTCA)} as a complementary objective for evaluating routing strategies in long-context distributed serving.
TTCA is the wall-clock time required to obtain the first correct answer.
This metric highlights a systems challenge: routers must account for both correctness and routing overhead when they rely on prompt characteristics.

Prior work explores semantic or content-aware routing that analyzes prompt intent or reasoning requirements with embedding-based classification and related prompt analysis techniques \cite{semantic_router_vllm2025}.
Semantic routing is well suited for environments where task heterogeneity is high and intent-level differentiation directly determines inference strategy (e.g., reasoning vs. non-reasoning paths, specialist vs. generalist models).
In contrast, our setting emphasizes stability under long-context accuracy variability.
In this setting, capability mismatches must be avoided and control-plane overhead must remain bounded.
Rather than replacing semantic routing, we explore complementary and lower-complexity designs.

We focus on retryable, task-oriented workloads in which a request has a stable target outcome and response correctness can be determined programmatically or via a reliable evaluation metric.
This assumption matches many existing LLM benchmarks and defines the scope in which TTCA is meaningful.
To isolate the system-level effect of accuracy-induced retries, our evaluation uses controlled key-value retrieval workloads that sweep context length and language.
Although synthetic, this workload provides a clean probe of context-length-dependent accuracy degradation that also appears in realistic settings such as document QA, summarization verification, and tool-based agents.

In this work, we introduce Lightweight Accuracy-Aware Routing (LAAR), a capability-based routing design for long-context workloads.
LAAR improves TTCA while keeping control-plane overhead bounded.
LAAR uses only lightweight prompt-derived features, such as prompt length and language, without invoking additional models or full-sequence semantic analysis.

This work makes the following contributions:

\begin{enumerate}[leftmargin=*, nosep]
\item \textbf{Characterizing long-context accuracy variability.}
We demonstrate that prompt length, language, and model selection amplify inference accuracy variance under long-context workloads.

\item \textbf{Revisiting routing objectives through TTCA.}
We introduce Time-to-Correct-Answer (TTCA) as a metric that captures retry-induced latency inflation, and argue that under long-context serving, accuracy becomes a system-level performance factor.

\item \textbf{Lightweight Accuracy-Aware Routing (LAAR).}
We introduce a capability-based routing design that stabilizes TTCA.
It reduces capability mismatches without relying on semantic intent inference.

\item \textbf{Empirical validation in long-context settings.}
We show that LAAR reduces mean TTCA in most evaluated settings across context lengths and languages.
It does so while keeping control-plane overhead bounded.
\end{enumerate}

\section{Background and Related Work}
\label{sec:motivation}

\subsection{Routing in Distributed LLM Serving}

Routing and scheduling algorithms are critical for lower latency and higher throughput in distributed LLM serving.
Performance-aware LLM load balancing~\cite{performanceaware2025} introduces heuristic- and RL-guided routing to adapt to workload characteristics.
Deferred Prefill~\cite{deferredprefill2025} mitigates decode-phase stalls by optimizing prompt departure timing.
ShapeShifter~\cite{shapeshifter2025} formulates LLM cluster management as a bin-packing problem and dynamically rebalances multi-dimensional resource utilization to improve GPU utilization.
SkyWalker~\cite{skywalker2025} extends routing to multi-region settings with prefix-aware traffic control.
Semantic routing approaches classify prompts by their reasoning requirements to guide inference-mode selection in vLLM-based stacks~\cite{semantic_router_vllm2025}.
Collectively, these systems treat routing as a resource allocation and load-balancing problem.
They optimize throughput, tail latency, or resource utilization.

However, existing routing policies assume that model correctness is stable once a request is assigned.
They therefore optimize for computational efficiency without modeling how correctness instability under long contexts can trigger retries and inflate user-visible delay.

\subsection{Long-Context Accuracy Variability}
\label{sec:long_context_variability}

A growing body of work systematically studies how LLM performance degrades under long contexts.
Lost in the Middle~\cite{lostinthemiddle2023} reveals strong positional bias and accuracy drops when relevant information appears in the middle of long inputs.
SCBench~\cite{li_scbench_2025} shows that long-context workloads often involve shared contexts, KV-cache reuse, and multi-round interactions that are missed by single-request evaluation.
RULER~\cite{ruler2024} demonstrates that the effective context length of models is often far below their advertised maximum, with task-dependent degradation.
LongBench~\cite{longbench2023} and Ada-LEval~\cite{adaleval2024} further report accuracy declines as input size grows, while MLNeedle~\cite{mlneedle2024} highlights language-dependent sensitivity in multilingual retrieval.
These prior studies establish that long-context accuracy is highly sensitive to position, task type, language, and model architecture.

However, most prior work evaluates long-context performance as a capability or benchmark problem rather than a serving-time systems problem.
They characterize degradation across positions, lengths, languages, or request patterns, but do not examine how accuracy variability across requests affects retry probability in real serving systems.
In distributed LLM serving, such variability can amplify user-visible delay when incorrect responses trigger retries.
Together, these gaps motivate our systems view: we treat long-context accuracy variability as a routing-time concern, quantify how it changes across models, languages, and context lengths, and evaluate routing with TTCA.

\section{Analysis of Model Accuracy and Latency}
\label{sec:accuracy_variability}

To understand how accuracy variability impacts retry behavior and routing decisions, we analyze single-shot model accuracy and latency under different context lengths and languages.
Prior sections focused on degradation within a single model.
Here, we emphasize cross-model variability, ranking changes, and their implications for multi-model serving.

\subsection{Single-shot experiments}

We run experiments with vLLM v0.16.0 on NVIDIA A100 GPUs (80GB VRAM).
Our dataset consists of 100 modified SCBench KV-lookup queries, split into two disjoint sets of 50.
This section reports single-shot results on the first split, which we also use to fit the offline estimators used by LAAR in Section~\ref{sec:laar}.
Every original query consists of a large context and a small question.
The contexts start with the prefix string ``JSON data: '' followed by a large JSON dictionary of random UUID key-value pairs.
The question is an English sentence that asks for the value associated with a key (e.g., ``Key: 6ab6ea3e-f288-4f33-ba46-7f42bb75b03f. The value associated with the specified key is:'').
We truncate the original contexts into 4K, 8K, 16K, 32K, and 64K tokens and translate the contexts and questions into Japanese and Chinese.
We run them sequentially on a single vLLM server that hosts Granite3.1-2B, Granite3.1-8B \cite{Granite-report}, Phi3-mini, Phi3-medium \cite{Phi3-report}, or Llama3.1-Swallow-8B \cite{fujii2024continualpretrainingcrosslingualllm,okazaki2024buildinglargejapaneseweb}.
We reuse the original SCBench script to identify correct answers and estimate accuracy for each model.

\begin{figure}[t]
    \centering
    \begin{subfigure}[t]{\columnwidth}
        \centering
        \includegraphics[width=\linewidth]{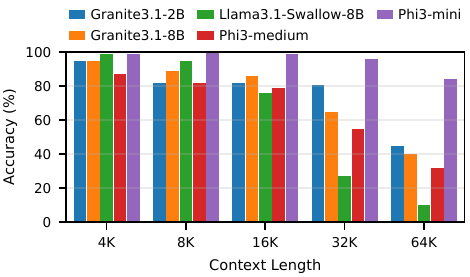}
        \caption{English prompts}
    \end{subfigure}

    \vspace{0.5em}

    \begin{subfigure}[t]{\columnwidth}
        \centering
        \includegraphics[width=\linewidth]{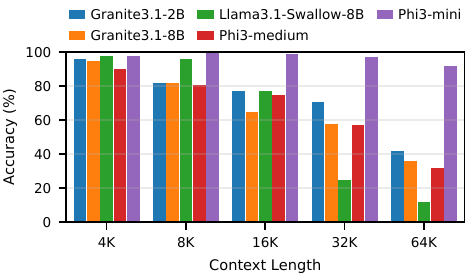}
        \caption{Japanese prompts}
    \end{subfigure}

    \vspace{0.5em}

    \begin{subfigure}[t]{\columnwidth}
        \centering
        \includegraphics[width=\linewidth]{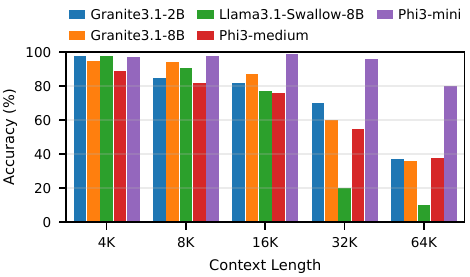}
        \caption{Chinese prompts}
    \end{subfigure}
    \caption{Mean accuracy of KV lookups with different context lengths and languages. Accuracy degradation is model- and language-dependent, and smaller models (e.g., Phi3-mini) can outperform larger ones.}
    \Description{Three stacked line charts compare mean KV-lookup accuracy across five models for English, Japanese, and Chinese prompts at context lengths from 4K to 64K tokens. Accuracy declines with longer contexts, but the decline differs by model and language.}
    \label{fig:minitest}
\end{figure}

Figure~\ref{fig:minitest} reports accuracy across models, languages, and context sizes.
Phi3-mini was often the most accurate across context lengths and notably outperformed Phi3-medium.
Granite3.1-2B underperformed Granite3.1-8B at smaller context lengths, but outperformed it at 32K and 64K.
Llama3.1-Swallow-8B exhibited a clear threshold-like failure:
it remained competitive up to 16K (and was often strong at 4K--16K), but collapsed sharply at 32K and further at 64K across languages.
We did not observe the single best model in terms of accuracy.
From a routing perspective, secondary model choices also matter when serving multiple queries concurrently in a cluster, but we still did not observe a consistent choice.
The parameter scale did not help predict accuracy in long-context KV lookups.

\begin{figure}[t]
    \centering
    \includegraphics[width=\linewidth]{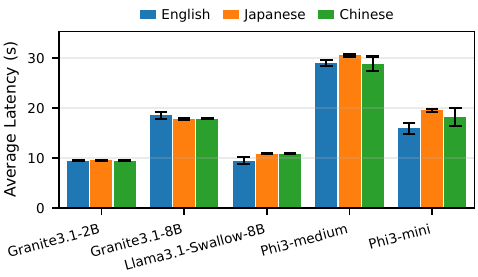}
    \caption{Mean latency for 64K contexts of KV lookups with five models. We omit other context sizes because they showed the same latency ranking among models.}
    \Description{A bar chart compares mean latency for 64K KV lookups across five models. Phi3-mini is slower than Granite3.1-2B and Llama3.1-Swallow-8B, while latency differences remain otherwise modest.}
    \label{fig:minitest_speed}
\end{figure}

Figure~\ref{fig:minitest_speed} shows the latency of each model with 64K tokens.
In contrast to the accuracy results, latency is consistent across context lengths and languages, but it is affected by the model.
Phi3-mini showed higher latency than Granite3.1-2B and Llama3.1-Swallow-8B.
The latency ranking is therefore more stable than the accuracy ranking.

\subsection{Routing Implications}

In summary, accuracy degradation slopes and effective context thresholds vary by model and language, whereas latency ranking is comparatively stable.
The best routing choice is therefore not stationary: model rankings change with prompt length and language, so routers must be context- and language-aware rather than fixed.
Model size does not monotonically predict long-context accuracy; smaller models such as Phi3-mini outperform larger 8B-class models in several ranges, particularly at mid-length contexts (8K--16K).
Language further amplifies this non-monotonicity, as a model that performs best in English is not necessarily optimal in Japanese or Chinese.
A policy that always selects the single highest-accuracy model can therefore be brittle and may concentrate load.
In practice, routers should prefer competitive models that jointly balance accuracy and latency, which motivates TTCA-aware routing.

\section{Time-to-Correct-Answer (TTCA)}
\label{sec:ttca}

From a user perspective, latency is not merely the time to receive an answer, but the time to receive a correct answer.
Under short-context workloads with stable accuracy, latency can often be approximated by a single inference time.
However, under long-context workloads, accuracy variability increases, and incorrect responses may trigger retries—either explicitly by users or implicitly by upstream systems.
In such cases, the effective delay experienced by users accumulates across attempts.

To capture this retry-induced latency inflation, we introduce Time-to-Correct-Answer (TTCA): the wall-clock time elapsed from the first attempt until the first correct answer is obtained.

For attempt $i$, let latency be $\ell_i$ and correctness be $C_i \in \{0,1\}$.
Correctness is determined by task-specific automatic evaluation (e.g., exact match in needle-in-a-haystack tasks).
If the first correct attempt is
\[
K = \min \{i \mid C_i = 1\},
\]
then TTCA is defined as
\[
\mathrm{TTCA} = \sum_{i=1}^{K} \ell_i.
\]

Intuitively, TTCA measures how long a user waits until they obtain a correct answer under retries.
In practice, retries are capped at $R$ attempts.
If no correct answer appears within $R$ attempts, the request is treated as failed and TTCA is right-censored at $\sum_{i=1}^{R}\ell_i$.

Unlike conventional per-request latency metrics, TTCA explicitly models retry dynamics induced by accuracy variability and therefore better reflects user-visible delay under long-context workloads.
We use TTCA as an evaluation objective rather than a production telemetry metric.

\section{Lightweight Accuracy-Aware Routing}
\label{sec:laar}

Lightweight Accuracy-Aware Routing (LAAR) ranks candidate models with a lightweight proxy for expected time-to-success.
The proxy combines estimates of success probability and serving latency under control-plane constraints.

\subsection{Routing Objective and Score}
\label{sec:schedule_objective}

Given a request $x$ and candidate model $m$, let $Q(m,x)$ denote the expected success probability and $L(m,x)$ the expected latency.
LAAR uses the following practical heuristic:

\[
cost(m \mid x) = \frac{L(m, x)}{Q(m, x)}.
\]

Under an idealized geometric retry model with stationary per-attempt success probability and latency, this score corresponds to expected time-to-success; in practice, LAAR uses $L(m, x)/Q(m, x)$ as a lightweight proxy for TTCA that is simple to evaluate online.
This approximation assumes independent retries with stationary success probability.

At routing time, LAAR computes this cost for each model and selects:
\[
m^* = \arg\min_m cost(m \mid x).
\]

This trade-off allows a slower model with higher success probability to outrank a faster but less reliable one.
We deliberately adopt simple estimators to keep routing efficient and scalable in the control plane.

To account for retry behavior, LAAR applies a penalty to models that have been previously selected for the same request.
This prevents repeated selection of the same model when its initial attempt fails.
It also encourages exploration of alternative models and improves time-to-success in multi-try scenarios.
This retry penalty is therefore a pragmatic extension beyond the idealized geometric interpretation above.
This is critical in practice because deterministic decoding can otherwise lead to repeated failures.

\subsection{Estimating Success Probability}
\label{sec:q_m_x}

To estimate the expected success probability $Q(m, x)$, LAAR uses a lightweight capability model that estimates the probability that a model produces a correct response given coarse, easily extractable request features.

Each request is mapped to a small set of features, such as language, task type, and input length bucket.
These features are inexpensive to extract and do not require deep semantic analysis.
In our current evaluation, we focus on a single task (KV lookup), so the task-type feature remains constant and is not explicitly used.

For each model, we train a logistic regression model offline that predicts the probability of success based on these features.
We choose logistic regression due to its simplicity, interpretability, and low inference overhead.
The resulting model is compact and can be evaluated efficiently at runtime.
In practice, this yields a function:

\[
Q(m, x) \in [0,1],
\]

which reflects the model's capability under the given request conditions.

This capability model serves as a static prior over model capabilities.
It complements the runtime latency estimation in Section~\ref{sec:schedule_objective}.
It is independent of runtime load and can therefore be precomputed and reused without introducing additional control-plane overhead.

\subsection{Estimating Latency}
\label{sec:l_m_x}

To estimate $L(m, x)$, we model latency as a function of the request size and the current load of the model.

Let $c(m)$ denote the empirical seconds per generated token for model $m$, obtained from offline measurements.
Let $T(x)$ denote the estimated number of tokens for request $x$, derived from the same length bucket used in Section~\ref{sec:q_m_x}.
We further denote by $R(m)$ the number of tokens currently being processed or waiting in the queue at model $m$.

The expected latency is then approximated as:

\[
L(m,x) = c(m) \cdot \big(T(x) + \alpha \cdot R(m)\big),
\]

where $\alpha$ is a constant that captures the impact of ongoing and queued work on latency. We use 0.7 for our experiments in Section~\ref{sec:evaluation}.

This formulation captures two key effects: (1) longer inputs incur higher processing cost, and (2) queueing and contention increase effective latency.
Because all inputs to this model are directly observable at runtime, it avoids expensive prediction pipelines and remains compatible with control-plane constraints.
This design deliberately avoids complex latency predictors.
It favors robustness and low overhead in the control plane over extra predictive accuracy.

\subsection{Implementation}

We implement LAAR as an Envoy \textit{Endpoint Picker (EPP)} policy via the external processing filter.
For a given request, our EPP computes a score for each candidate endpoint that serves models.
It then uses llm-d's \texttt{MaxScorePicker} \cite{gateway-api-inference-extension} to forward the request to the endpoint with the maximum score.
At request time, Envoy invokes our extension.
The extension extracts lightweight request features, evaluates the success probability $Q(m, x)$ and the expected latency $L(m, x)$ for each model, and selects the model that minimizes the cost $cost(m \mid x)$, i.e., maximizes the inverted cost for MaxScorePicker.

Our routing logic relies on lightweight CPU-side computations within the request processing path.
Feature extraction parses a short sampled substring of the user text.
Language is inferred from character classes (ASCII vs. CJK and Hiragana/Katakana).
Thus, the core decision is realized as a compact if-else cascade to avoid semantic parsing or auxiliary model inference.

All computations are constant-time per target model.
Where $\mathcal{M}$ is the set of candidate models evaluated at routing time, the overall complexity is therefore $O(|\mathcal{M}|)$.
The router relies only on locally available information and does not require cross-backend coordination or global state.
This design satisfies control-plane boundedness: routing decisions incur minimal overhead and do not become a bottleneck even under high request rates.

To support retry-aware routing, the router requires information about previously selected models for the same request.
We propagate this information via request metadata: the router returns the selected model identifier to the client, which includes it in subsequent retry requests.
This enables the router to apply penalties to previously attempted models without any server-side session state.

\section{Preliminary Evaluation}
\label{sec:evaluation}

Our primary goal in this evaluation is to examine whether LAAR improves TTCA under long-context workloads.
For comparison, we evaluate two representative baselines, load-aware routing and session-affinity routing, by directly invoking the llm-d implementation \cite{llmd-inference-scheduler}.
We replace only the scoring logic to keep the gateway and forwarding path identical among all the experiments.

\subsection{Experimental Setup}

We evaluate a distributed serving stack composed of an Envoy-based EPP router and a pool of vLLM instances.
Our primary metric is TTCA (Section~\ref{sec:ttca}), with up to ten request retries, i.e., a retry cap \(R=10\).

All experiments run vLLM v0.16.0 on NVIDIA A100 GPUs (80GB VRAM) interconnected via a 10Gbps network.
The cluster contains five vLLM instances, which run the five models described in Section~\ref{sec:accuracy_variability}.
The number of concurrent requests is set to eight.
We use deterministic decoding (temperature \(=0\)) to reduce output variance and isolate routing effects.
Workloads consist of SCBench key-value lookups with different context sizes and languages, as shown in Section~\ref{sec:accuracy_variability}.
To eliminate cross-request KV-cache reuse effects, we carefully iterate over different query sets.

This evaluation uses only the held-out second 50-query split.
The first split from Section~\ref{sec:accuracy_variability} is used offline to fit the success probability model $Q(m, x)$ and estimate latency $L(m, x)$ as described in Section~\ref{sec:laar}.

\subsection{TTCA and Success Rate}

\begin{figure}[t]
    \centering
    \includegraphics[width=\linewidth]{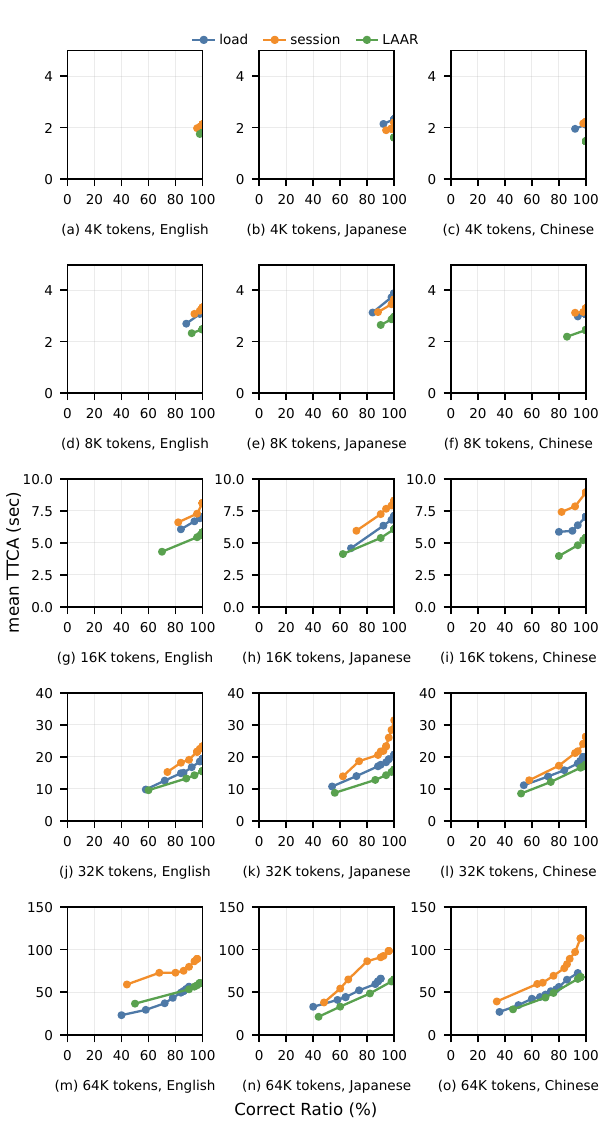}
    \caption{TTCA and success rate for retryable UUID key-value lookups in 4K, 8K, 16K, 32K, and 64K contexts under English, Japanese, and Chinese with load-aware, session-affinity, and LAAR routing. Retries monotonically increase both TTCA and the success rate, and we run up to ten retries. However, LAAR finishes within at most five attempts because it avoids reusing failed models.}
    \Description{A grid of plots compares TTCA and success rate across retry counts for load-aware routing, session-affinity routing, and LAAR over multiple context lengths and languages. LAAR reaches higher final success rates with fewer distinct attempts, while TTCA rises as retries accumulate.}
    \label{fig:ttca}
\end{figure}

Figure~\ref{fig:ttca} shows the relationship between TTCA and success rate across retries for different routing policies.
Overall, retries allowed the system to improve the success rate at the cost of increased TTCA, i.e., retry-induced latency.
As illustrated in the figure, regardless of routing strategy or language, shorter contexts consistently achieved higher success rates, while success rates degraded as context length increased.

LAAR sometimes started with a lower first-attempt success rate than the other methods, yet achieved lower TTCA over the retry process.
This indicates that TTCA depends on the sequence of model choices and their latency, not only on first-attempt success in isolation.
As retries proceeded, LAAR's success rate steadily improved and reached the highest final success rate among all methods.

In contrast, both session-affinity routing and load-aware routing benefited less from additional retries than LAAR.
These approaches often routed repeated attempts to the same model, yielding prefix reuse but limited exploration of alternatives.
By explicitly avoiding previously failed models, LAAR was able to cover a broader set of candidate models within a small number of retries, which led to higher final success rates in most settings, although some 64K cases remained unsolved by all models.

\begin{figure}[t]
    \centering
    \includegraphics[width=\linewidth]{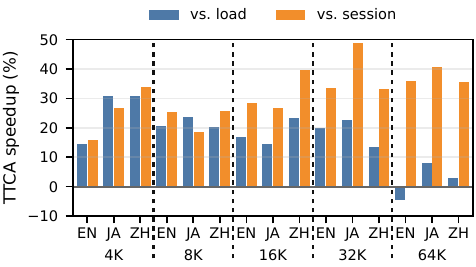}
    \caption{TTCA improvement ratio for UUID key-value lookups in 4K, 8K, 16K, 32K, and 64K contexts under English (EN), Japanese (JA), and Chinese (ZH) compared to load-aware and session-affinity routing.}
    \Description{A grouped bar chart shows LAAR's TTCA improvement ratios over load-aware and session-affinity routing for different context lengths and languages. Improvements are generally positive and are larger than those over session-affinity routing at longer contexts.}
    \label{fig:improvement}
\end{figure}

Figure~\ref{fig:improvement} presents the relative TTCA improvement of LAAR at the final attempt (up to 10 retries).
As shown in the figure, LAAR achieved up to 31\% improvement over load-aware routing and 49\% over session-affinity routing.

For load-aware routing, performance became relatively closer to LAAR as context length increased.
This was because larger contexts increased overall system load.
Load balancing then improved latency, which directly reduced TTCA.
In the 64K-token case, load-aware routing even achieved lower absolute TTCA than LAAR.
This highlights that latency-dominant regimes can favor load balancing.
However, as observed in Figure~\ref{fig:ttca}, its final success rate was lower.
This result reflects the trade-off between lower latency and lower correctness.

In contrast, session-affinity routing degraded relative to LAAR as context length increased.
This is primarily due to its policy of consistently routing requests within the same session to the same model.
That policy becomes suboptimal when model accuracy varies with context length.
While all models exhibit high accuracy for short contexts, longer contexts introduce divergence in model performance.
That divergence amplifies the advantage of LAAR's adaptive routing and leads to larger improvements.

\section{Discussion and Future Directions}

The experimental results in Section \ref{sec:evaluation} highlight that TTCA is a meaningful objective for routing and scheduling in LLM serving systems.
At the same time, they suggest that the LAAR algorithm may require further adaptation depending on factors such as available models, task types, context length, and language.

Although LAAR introduces additional computation, its overhead remains sufficiently small compared to model inference latency (e.g., on the order of milliseconds even for 64K-token inputs).
While this overhead is not explicitly included in TTCA, it may need to be incorporated into the latency term $L(m, x)$ when extending the approach to more complex routing logic.

Finally, we evaluate load-aware and session-affinity routing as baselines, but hybrid approaches that combine them with LAAR may further improve performance.
For example, the relatively low latency of load-aware routing in the 64K-token setting suggests that integrating load-awareness into LAAR could yield better TTCA under high-load conditions.
Session-affinity may provide benefits in workloads not considered here, such as multi-turn interactions or heterogeneous task sequences.
Prefix locality reduces prefill cost, but strict stickiness may repeatedly query an incompatible model under deterministic decoding and hurt TTCA.
Conversely, aggressively switching to accuracy-compatible endpoints may sacrifice cache reuse.
Designing a multi-objective router that jointly optimizes correctness and cache locality remains future work.

Although we focus on single-intent, retryable workloads, TTCA can be extended to more general conversational settings.
For example, TTCA can be applied at the level of individual turns, or more broadly to measure time-to-goal under a higher-level notion of conversational success.
An important direction for future work is to define such success signals, potentially through proxy metrics such as LLM-as-a-judge or human feedback.

\section{Conclusion}

Long-context LLM serving brings routing quality and correctness into closer interaction because a poor model choice can trigger expensive retries.
TTCA makes this coupling explicit by measuring time-to-correct-answer rather than single-shot latency.

From this perspective, we introduced LAAR, a lightweight routing design for routing across heterogeneous models.
LAAR combines success-probability and latency estimates.
In our preliminary key-value retrieval evaluation, LAAR improved TTCA without semantic routing or auxiliary model inference in the control plane.
Extending the approach to broader open-ended tasks (e.g., \texttt{repoqa}, \texttt{summary}) remains future work because correctness criteria become more task-specific.

\bibliographystyle{ACM-Reference-Format}
\bibliography{refs}

@inproceedings{vllm_sosp23,
author = {Kwon, Woosuk and Li, Zhuohan and Zhuang, Siyuan and Sheng, Ying and Zheng, Lianmin and Yu, Cody Hao and Gonzalez, Joseph and Zhang, Hao and Stoica, Ion},
title = {Efficient Memory Management for Large Language Model Serving with PagedAttention},
year = {2023},
isbn = {9798400702297},
publisher = {Association for Computing Machinery},
address = {New York, NY, USA},
url = {https://doi.org/10.1145/3600006.3613165},
doi = {10.1145/3600006.3613165},
booktitle = {Proceedings of the 29th Symposium on Operating Systems Principles},
pages = {611–626},
numpages = {16},
location = {Koblenz, Germany},
series = {SOSP '23}
}

@misc{li_scbench_2025,
      title={SCBench: A KV Cache-Centric Analysis of Long-Context Methods}, 
      author={Yucheng Li and Huiqiang Jiang and Qianhui Wu and Xufang Luo and Surin Ahn and Chengruidong Zhang and Amir H. Abdi and Dongsheng Li and Jianfeng Gao and Yuqing Yang and Lili Qiu},
      year={2025},
      eprint={2412.10319},
      archivePrefix={arXiv},
      primaryClass={cs.CL},
      url={https://arxiv.org/abs/2412.10319}, 
}

@misc{skywalker2025,
      title={SkyWalker: A Locality-Aware Cross-Region Load Balancer for LLM Inference}, 
      author={Tian Xia and Ziming Mao and Jamison Kerney and Ethan J. Jackson and Zhifei Li and Jiarong Xing and Scott Shenker and Ion Stoica},
      year={2025},
      eprint={2505.24095},
      archivePrefix={arXiv},
      primaryClass={cs.DC},
      url={https://arxiv.org/abs/2505.24095}, 
}

@misc{semantic_router_vllm2025,
      title={When to Reason: Semantic Router for vLLM}, 
      author={Chen Wang and Xunzhuo Liu and Yuhan Liu and Yue Zhu and Xiangxi Mo and Junchen Jiang and Huamin Chen},
      year={2025},
      eprint={2510.08731},
      archivePrefix={arXiv},
      primaryClass={cs.ET},
      url={https://arxiv.org/abs/2510.08731}, 
}

@inproceedings{performanceaware2025,
author = {Jain, Kunal and Parayil, Anjaly and Mallick, Ankur and Choukse, Esha and Qin, Xiaoting and Zhang, Jue and Goiri, \'{I}\~{n}igo and Wang, Rujia and Bansal, Chetan and R\"{u}hle, Victor and Kulkarni, Anoop and Kofsky, Steve and Rajmohan, Saravan},
title = {Performance Aware LLM Load Balancer for Mixed Workloads},
year = {2025},
isbn = {9798400715389},
publisher = {Association for Computing Machinery},
address = {New York, NY, USA},
url = {https://doi.org/10.1145/3721146.3721947},
doi = {10.1145/3721146.3721947},
booktitle = {Proceedings of the 5th Workshop on Machine Learning and Systems},
pages = {19–30},
numpages = {12},
location = {World Trade Center, Rotterdam, Netherlands},
series = {EuroMLSys '25}
}

@inproceedings{deferredprefill2025,
author = {Mohanty, Moonmoon and Bolar, Gautham and Patil, Preetam and Devi, UmaMaheswari and George, Felix and Moogi, Pratibha and Parag, Parimal},
title = {Deferred prefill for throughput maximization in LLM inference},
year = {2025},
isbn = {9798400715389},
publisher = {Association for Computing Machinery},
address = {New York, NY, USA},
url = {https://doi.org/10.1145/3721146.3721962},
doi = {10.1145/3721146.3721962},
booktitle = {Proceedings of the 5th Workshop on Machine Learning and Systems},
pages = {100–106},
numpages = {7},
location = {World Trade Center, Rotterdam, Netherlands},
series = {EuroMLSys '25}
}

@inproceedings{shapeshifter2025,
author = {Lai, Ruiqi and Cao, Siyu and Li, Leqi and Mai, Luo and Ustiugov, Dmitrii},
title = {Manage the Workloads not the Cluster: Designing a Control Plane for Large-Scale AI Clusters},
year = {2025},
isbn = {9798400715389},
publisher = {Association for Computing Machinery},
address = {New York, NY, USA},
url = {https://doi.org/10.1145/3721146.3721937},
doi = {10.1145/3721146.3721937},
booktitle = {Proceedings of the 5th Workshop on Machine Learning and Systems},
pages = {246–253},
numpages = {8},
location = {World Trade Center, Rotterdam, Netherlands},
series = {EuroMLSys '25}
}

@misc{contextparallel2025,
      title={Context Parallelism for Scalable Million-Token Inference}, 
      author={Amy Yang and Jingyi Yang and Aya Ibrahim and Xinfeng Xie and Bangsheng Tang and Grigory Sizov and Jeremy Reizenstein and Jongsoo Park and Jianyu Huang},
      year={2025},
      eprint={2411.01783},
      archivePrefix={arXiv},
      primaryClass={cs.DC},
      url={https://arxiv.org/abs/2411.01783}, 
}

@misc{llmd-inference-scheduler,
  author = {{llm-d Project}},
  title  = {llm-d Inference Scheduler},
  year   = {2026},
  url    = {https://github.com/llm-d/llm-d-inference-scheduler},
  note   = {GitHub repository. Accessed: 2026-04-10}
}

@misc{gateway-api-inference-extension,
  author = {{Kubernetes SIG Network}},
  title  = {Gateway API Inference Extension},
  year   = {2026},
  url    = {https://github.com/kubernetes-sigs/gateway-api-inference-extension},
  note   = {GitHub repository. Accessed: 2026-04-10}
}

@misc{lostinthemiddle2023,
      title={Lost in the Middle: How Language Models Use Long Contexts}, 
      author={Nelson F. Liu and Kevin Lin and John Hewitt and Ashwin Paranjape and Michele Bevilacqua and Fabio Petroni and Percy Liang},
      year={2023},
      eprint={2307.03172},
      archivePrefix={arXiv},
      primaryClass={cs.CL},
      url={https://arxiv.org/abs/2307.03172}, 
}

@misc{ruler2024,
      title={RULER: What's the Real Context Size of Your Long-Context Language Models?}, 
      author={Cheng-Ping Hsieh and Simeng Sun and Samuel Kriman and Shantanu Acharya and Dima Rekesh and Fei Jia and Yang Zhang and Boris Ginsburg},
      year={2024},
      eprint={2404.06654},
      archivePrefix={arXiv},
      primaryClass={cs.CL},
      url={https://arxiv.org/abs/2404.06654}, 
}

@misc{longbench2023,
      title={LongBench: A Bilingual, Multitask Benchmark for Long Context Understanding}, 
      author={Yushi Bai and Xin Lv and Jiajie Zhang and Hongchang Lyu and Jiankai Tang and Zhidian Huang and Zhengxiao Du and Xiao Liu and Aohan Zeng and Lei Hou and Yuxiao Dong and Jie Tang and Juanzi Li},
      year={2024},
      eprint={2308.14508},
      archivePrefix={arXiv},
      primaryClass={cs.CL},
      url={https://arxiv.org/abs/2308.14508}, 
}

@misc{adaleval2024,
      title={Ada-LEval: Evaluating long-context LLMs with length-adaptable benchmarks}, 
      author={Chonghua Wang and Haodong Duan and Songyang Zhang and Dahua Lin and Kai Chen},
      year={2024},
      eprint={2404.06480},
      archivePrefix={arXiv},
      primaryClass={cs.CL},
      url={https://arxiv.org/abs/2404.06480}, 
}

@misc{mlneedle2024,
      title={Multilingual Needle in a Haystack: Investigating Long-Context Behavior of Multilingual Large Language Models}, 
      author={Amey Hengle and Prasoon Bajpai and Soham Dan and Tanmoy Chakraborty},
      year={2024},
      eprint={2408.10151},
      archivePrefix={arXiv},
      primaryClass={cs.CL},
      url={https://arxiv.org/abs/2408.10151}, 
}

@misc{Phi3-report,
      title={Phi-3 Technical Report: A Highly Capable Language Model Locally on Your Phone}, 
      author={Marah Abdin and Jyoti Aneja and Hany Awadalla and Ahmed Awadallah and Ammar Ahmad Awan and Nguyen Bach and Amit Bahree and Arash Bakhtiari and Jianmin Bao and Harkirat Behl and Alon Benhaim and Misha Bilenko and Johan Bjorck and Sébastien Bubeck and Martin Cai and Qin Cai and Vishrav Chaudhary and Dong Chen and Dongdong Chen and Weizhu Chen and Yen-Chun Chen and Yi-Ling Chen and Hao Cheng and Parul Chopra and Xiyang Dai and Matthew Dixon and Ronen Eldan and Victor Fragoso and Jianfeng Gao and Mei Gao and Min Gao and Amit Garg and Allie Del Giorno and Abhishek Goswami and Suriya Gunasekar and Emman Haider and Junheng Hao and Russell J. Hewett and Wenxiang Hu and Jamie Huynh and Dan Iter and Sam Ade Jacobs and Mojan Javaheripi and Xin Jin and Nikos Karampatziakis and Piero Kauffmann and Mahoud Khademi and Dongwoo Kim and Young Jin Kim and Lev Kurilenko and James R. Lee and Yin Tat Lee and Yuanzhi Li and Yunsheng Li and Chen Liang and Lars Liden and Xihui Lin and Zeqi Lin and Ce Liu and Liyuan Liu and Mengchen Liu and Weishung Liu and Xiaodong Liu and Chong Luo and Piyush Madan and Ali Mahmoudzadeh and David Majercak and Matt Mazzola and Caio César Teodoro Mendes and Arindam Mitra and Hardik Modi and Anh Nguyen and Brandon Norick and Barun Patra and Daniel Perez-Becker and Thomas Portet and Reid Pryzant and Heyang Qin and Marko Radmilac and Liliang Ren and Gustavo de Rosa and Corby Rosset and Sambudha Roy and Olatunji Ruwase and Olli Saarikivi and Amin Saied and Adil Salim and Michael Santacroce and Shital Shah and Ning Shang and Hiteshi Sharma and Yelong Shen and Swadheen Shukla and Xia Song and Masahiro Tanaka and Andrea Tupini and Praneetha Vaddamanu and Chunyu Wang and Guanhua Wang and Lijuan Wang and Shuohang Wang and Xin Wang and Yu Wang and Rachel Ward and Wen Wen and Philipp Witte and Haiping Wu and Xiaoxia Wu and Michael Wyatt and Bin Xiao and Can Xu and Jiahang Xu and Weijian Xu and Jilong Xue and Sonali Yadav and Fan Yang and Jianwei Yang and Yifan Yang and Ziyi Yang and Donghan Yu and Lu Yuan and Chenruidong Zhang and Cyril Zhang and Jianwen Zhang and Li Lyna Zhang and Yi Zhang and Yue Zhang and Yunan Zhang and Xiren Zhou},
      year={2024},
      eprint={2404.14219},
      archivePrefix={arXiv},
      primaryClass={cs.CL},
      url={https://arxiv.org/abs/2404.14219}, 
}

@misc{fujii2024continualpretrainingcrosslingualllm,
      title={Continual Pre-Training for Cross-Lingual LLM Adaptation: Enhancing Japanese Language Capabilities}, 
      author={Kazuki Fujii and Taishi Nakamura and Mengsay Loem and Hiroki Iida and Masanari Ohi and Kakeru Hattori and Hirai Shota and Sakae Mizuki and Rio Yokota and Naoaki Okazaki},
      year={2024},
      eprint={2404.17790},
      archivePrefix={arXiv},
      primaryClass={cs.CL},
      url={https://arxiv.org/abs/2404.17790}, 
}

@misc{okazaki2024buildinglargejapaneseweb,
      title={Building a Large Japanese Web Corpus for Large Language Models}, 
      author={Naoaki Okazaki and Kakeru Hattori and Hirai Shota and Hiroki Iida and Masanari Ohi and Kazuki Fujii and Taishi Nakamura and Mengsay Loem and Rio Yokota and Sakae Mizuki},
      year={2024},
      eprint={2404.17733},
      archivePrefix={arXiv},
      primaryClass={cs.CL},
      url={https://arxiv.org/abs/2404.17733}, 
}

@misc{Granite-report,
    author={{IBM Granite Team}},
    year=2024,
    title={Granite 3.0 language models},
}

@misc{jin2024longcontextllmsmeetrag,
      title={Long-Context LLMs Meet RAG: Overcoming Challenges for Long Inputs in RAG}, 
      author={Bowen Jin and Jinsung Yoon and Jiawei Han and Sercan O. Arik},
      year={2024},
      eprint={2410.05983},
      archivePrefix={arXiv},
      primaryClass={cs.CL},
      url={https://arxiv.org/abs/2410.05983}, 
}

@misc{xu2026evolutiontoolusellm,
      title={The Evolution of Tool Use in LLM Agents: From Single-Tool Call to Multi-Tool Orchestration}, 
      author={Haoyuan Xu and Chang Li and Xinyan Ma and Xianhao Ou and Zihan Zhang and Tao He and Xiangyu Liu and Zixiang Wang and Jiafeng Liang and Zheng Chu and Runxuan Liu and Rongchuan Mu and Dandan Tu and Ming Liu and Bing Qin},
      year={2026},
      eprint={2603.22862},
      archivePrefix={arXiv},
      primaryClass={cs.SE},
      url={https://arxiv.org/abs/2603.22862}, 
}

@inproceedings{10.5555/3737916.3740957,
author = {Ma, Yubo and Zang, Yuhang and Chen, Liangyu and Chen, Meiqi and Jiao, Yizhu and Li, Xinze and Lu, Xinyuan and Liu, Ziyu and Ma, Yan and Dong, Xiaoyi and Zhang, Pan and Pan, Liangming and Jiang, Yu-Gang and Wang, Jiaqi and Cao, Yixin and Sun, Aixin},
title = {MMLONGBENCH-DOC: benchmarking long-context document understanding with visualizations},
year = {2024},
isbn = {9798331314385},
publisher = {Curran Associates Inc.},
address = {Red Hook, NY, USA},
booktitle = {Proceedings of the 38th International Conference on Neural Information Processing Systems},
articleno = {3041},
numpages = {48},
location = {Vancouver, BC, Canada},
series = {NIPS '24}
}

@misc{yuan2026dualmap,
      title={DualMap: Enabling Both Cache Affinity and Load Balancing for Distributed LLM Serving}, 
      author={Ying Yuan and Pengfei Zuo and Bo Wang and Zhangyu Chen and Zhipeng Tan and Zhou Yu},
      year={2026},
      eprint={2602.06502},
      archivePrefix={arXiv},
      primaryClass={cs.DC},
      url={https://arxiv.org/abs/2602.06502}, 
}

@INPROCEEDINGS {11120553,
author = { Recasens, Pol G. and Agullo, Ferran and Zhu, Yue and Wang, Chen and Lee, Eun Kyung and Tardieu, Olivier and Torres, Jordi and Berral, Josep Ll. },
booktitle = { 2025 IEEE 18th International Conference on Cloud Computing (CLOUD) },
title = {{ Mind the Memory Gap: Unveiling GPU Bottlenecks in Large-Batch LLM Inference }},
year = {2025},
volume = {},
ISSN = {},
pages = {277-287},
doi = {10.1109/CLOUD67622.2025.00036},
url = {https://doi.ieeecomputersociety.org/10.1109/CLOUD67622.2025.00036},
publisher = {IEEE Computer Society},
address = {Los Alamitos, CA, USA},
month =Jul}

@misc{yang2025lserveefficientlongsequencellm,
      title={LServe: Efficient Long-sequence LLM Serving with Unified Sparse Attention}, 
      author={Shang Yang and Junxian Guo and Haotian Tang and Qinghao Hu and Guangxuan Xiao and Jiaming Tang and Yujun Lin and Zhijian Liu and Yao Lu and Song Han},
      year={2025},
      eprint={2502.14866},
      archivePrefix={arXiv},
      primaryClass={cs.CL},
      url={https://arxiv.org/abs/2502.14866}, 
}

@misc{haseeb2025contextengineeringmultiagentllm,
      title={Context Engineering for Multi-Agent LLM Code Assistants Using Elicit, NotebookLM, ChatGPT, and Claude Code}, 
      author={Muhammad Haseeb},
      year={2025},
      eprint={2508.08322},
      archivePrefix={arXiv},
      primaryClass={cs.SE},
      url={https://arxiv.org/abs/2508.08322}, 
}
\end{document}